\documentclass[aps,prl,twocolumn,superscriptaddress,floatfix,a4]{revtex4}
\usepackage{epsfig,amsmath,amssymb,color}
\begin{document}

\title{Pair correlations of a spin-imbalanced Fermi gas on two-leg ladders}
\author{A. E. Feiguin} 
\affiliation{Microsoft Project Q, University of California,  Santa Barbara,  CA 93106, USA}
\affiliation{Condensed Matter Theory Center,  Department of Physics, University of Maryland, College Park, MD 20742, USA}
\author{F. Heidrich-Meisner}
\affiliation{Institut f\"ur Theoretische Physik C, RWTH Aachen University, 52056 Aachen}

\date{Feb. 4, 2009}

\begin{abstract}
We study the pair correlations of a spin-imbalanced two-leg ladder with attractive interactions, 
using the density matrix renormalization group method (DMRG). We identify regions in the phase diagram 
spanned by the chemical potential and the magnetic field that can harbor Fulde-Ferrell-Larkin-Ovchinnikov (FFLO)-like physics. 
Results for the pair structure factor, exhibiting multiple pairing wave-vectors, 
substantiate the presence of FFLO-like correlations.  We further discuss phase separation scenarios induced by a harmonic trap, 
which differ from the case of isolated chains.
\end{abstract}

\maketitle

The experimental realization of fermionic superfluids
in ultracold atom gases
under clean conditions and with a great control over interactions 
has paved the way toward a detailed understanding of the BEC-BCS crossover
of spin-balanced, ultracold Fermi gases \cite{reviews}. 
Now, the case of a polarized two-component Fermi gas, realized by unequally populating the two lowest hyperfine states, 
has moved into the focus of current experimental work \cite{exp}.

Intriguing properties such as phase separation in a trap and the transition from superfluidity
to a normal state driven by the population imbalance have attracted a great deal of 
attention, but it is, in particular, the search for exotic superfluids  such as the
FFLO one \cite{fulde64,larkin64} that drives the current interest in imbalanced Fermi gases.
In an FFLO state, the order parameter is spatially inhomogeneous with Cooper pairs with a finite
center-of-mass momentum. In recent experiments on three dimensional (3D) ultracold gases, this state
remains elusive, and theoretical work indicates that in 3D, the phase space volume of this phase
in the interaction-polarization plane is small \cite{3D-imbalanced}.
Yet, reducing the spatial dimension  renders this pairing mechanism  more
effective as a larger portion of the Fermi surfaces of minority and majority spins can be matched 
\cite{parish07}.
In particular, in one dimension (1D), where a true condensation is prohibited, the existence of an 
FFLO-type state with quasi-long range order has been proven by means of analytical
\cite{yang01,hu07} as well as
numerically exact approaches \cite{roux,feiguin07,fflo_numerical,luescher08}.  This also pertains to the experimentally relevant case of a harmonic trap 
\cite{feiguin07,fflo_numerical,luescher08}. 

It is then natural to ask whether such quasi-FFLO states seen in 1D are stable against
coupling chains to 2D or 3D ensembles, in order to connect the aforementioned theoretical results for 1D  to those
available for 2D \cite{parish07,moreo07,koponen08,iskin08,he08}. For 1D chains weakly coupled to a 3D
array, the polarization-interaction phase diagram   has recently been derived in Refs.~\cite{yang01,parish07,zhao08,luescher08}. Here we present a rigorous and quasi-exact numerical  analysis of pairing
correlations in two coupled chains, using  DMRG \cite{dmrg}. 

The physics of spin \cite{dagotto} and Hubbard ladders
\cite{giamarchi} has proven to be 
unique and interesting  in itself, due to the emergence of exotic quantum phases -- such as spin liquids -- driven by strong correlations \cite{dagotto}.
The experimentally realization of ladders in optical lattices  
as arrays of double wells 
could be an important step toward understanding the experimental and theoretical challenges 
that we may face before scaling to 2D systems. 
Double wells have already been studied in recent experiments \cite{lee07} and Danshita {\it et al.} \cite{danshita08} 
have given a prescription 
 of how the parameters of a double well potential need to be tuned to create a ladder system in an optical lattice.
Methods to detect the FFLO state in experiments using noise correlations
or interferometry are discussed in Refs.~\cite{luescher08,gritsev08}.

We study the  
Hubbard model on a $2\times L$ ladder:
\begin{eqnarray}
H &=& - t_{\perp}\sum_{i=1,\sigma}^L  (c_{i,1,\sigma}^{\dagger}   c_{i,2,\sigma} + h.c.)+ U \sum_{l,i=1,\sigma}^{L} n_{i,l,\uparrow} n_{i,l,\downarrow}  \nonumber \\
  &&  -t_{\parallel} \sum_{l,i=1,\sigma}^{L-1} (c_{i,l,\sigma}^{\dagger}   c_{i+1,l,\sigma} + h.c)  -\mu N - h N p/2\nonumber\\
  && -V_{\mathrm{trap}} \sum_{l,i=1,\sigma}^L \lbrack i-(L+1)/2\rbrack^2 n_{i,l,\sigma} \label{eq:ham}\,,
\end{eqnarray}
where $c_{i,l,\sigma}^{(\dagger)}$ is a fermionic annihilation (creation) operator acting on a site on rung $i$ and leg $l$ ($l=1,2$).
The hopping matrix elements along rungs and legs are denoted by $t_{\perp}$ and $t_{\parallel}$, respectively.
We study the case of strongly attractive onsite interactions $U=-8t_{\parallel}$ 
(see 
Ref.~\cite{danshita08} for a discussion of how  $U$, $t_{\parallel}$,
and $t_{\perp}$ are related to experimentally controllable parameters in an optical lattice).
Further, $n_{i,l,\sigma}=c_{i,l,\sigma}^{\dagger}c_{i,l,\sigma}^{}$, yielding the number of fermions of each species as $N_{\sigma}=\sum_{l,i} \langle n_{i,l,\sigma} \rangle$,
with $N=N_{\uparrow}+N_{\downarrow}$ and the pseudo-spin index $\sigma=\uparrow,\downarrow$.
As customary, $h$ is the magnetic field, $\mu$ the chemical potential,
$n=N/(2L)$  the filling factor, and $p=(N_{\uparrow}-N_{\downarrow})/N$ measures the polarization (we use $N_{\uparrow}>N_{\downarrow}$). A harmonic trapping potential 
is introduced by the last term in Eq.~(\ref{eq:ham}).

Our analysis is mostly concerned with the case of $t_{\perp} = t_{\parallel}$
and thus not restricted to weakly coupled chains.
We first 
study the emergence of FFLO-like correlations as a function of 
the ratio $t_{\perp}/t_{\parallel}$ and filling $n$ for  $V_{\mathrm{trap}}=0$.

\begin{figure}[t]
\includegraphics[width=0.23\textwidth,angle=-90]{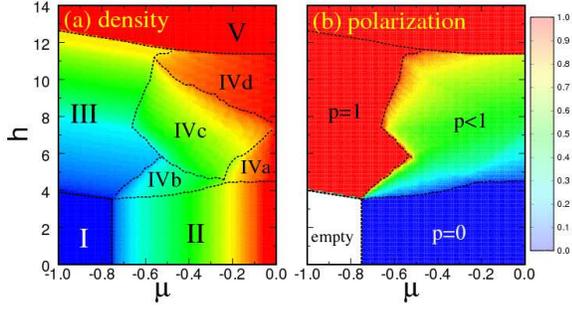}
\caption{(Color online) Bulk phase diagrams ($U=-8t_{\parallel}$,
$t_{\perp}=t_{\parallel}$) in the magnetic field $h$ 
vs. chemical potential $\mu$ plane ($L=30$ data): (a): filling $n$; (b) polarization $p$ (see the legend for the color coding; dotted lines: guide to the eye).
The phases I-V are defined in the text. Unlike the $U>0$ case \cite{roux}, no plateau at $p\propto $ hole-concentration is found.
}
\label{fig:phase}
\end{figure}

The bulk phase diagram  is displayed in 
Figs.~\ref{fig:phase} (a) and (b), showing contour plots of $n$  and $p$ in the  $h$ vs. $\mu$ plane, respectively.
We infer the presence of five phases
(for the single chain case, see Ref.~\cite{essler-book}):
I: the empty phase ($n=0$, $p=0$);
 II: a spin gapped phase ($p=0$, $n\leq 1$);
 III: fully polarized, but less than half filled ($p=n$, $n<1$); 
 IV: partially polarized ($n<1$);
 V: fully polarized ($n=1$).
Phase IV is the candidate for FFLO-like physics and we will
show that a rich structure in the real-space modulation of the pair correlation function emerges.

\begin{figure}[t]
\includegraphics[width=0.44\textwidth]{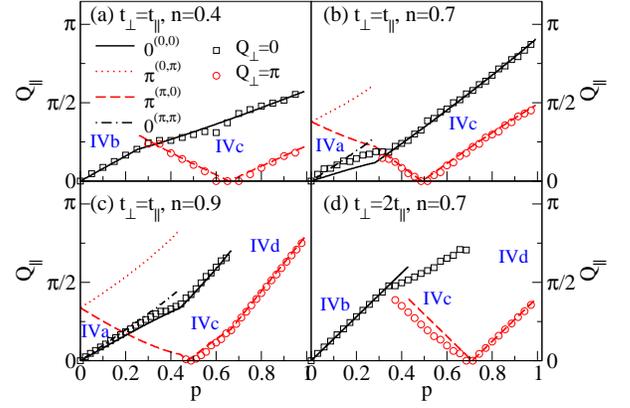}
\caption{(Color online)
Pairing wave-vectors $Q_{\parallel}$ extracted from   (i) 
$Q_{\parallel}=k_{F\parallel}^{\uparrow}-k_{F\parallel}^{\downarrow}$  
 at $U=0$  (lines) and (ii) the position of the maximum in the pair structure factor (DMRG  for  $L=50$ at $U=-8t_{\parallel}$; squares: $Q_{\perp}=0$; circles: $Q_{\perp}=\pi$).  
(a)--(c) $t_{\parallel}=t_{\perp}$ with $n=0.4, 0.7$, and $0.9$, respectively.
(d)  $t_{\perp}=2t_{\parallel}$, $n=0.7$. 
The lines [see the legend in (a)] represent the possible branches (see the text for details).}
\label{fig:Q}
\end{figure}

As a guidance for the interpretation of our DMRG results, we next
consider the noninteracting limit (see, {\it e.g.}, Ref.~\cite{roux}). The dispersion relation at $U=0$
with periodic boundary conditions along the legs of the ladder is:
$\epsilon^{\sigma}(k_{\parallel}) = -2t_{\parallel} \cos(k_{\parallel}) -t_{\perp} \cos(k_{\perp}^{\sigma})$, {\it i.e.}, there are four bands
which we label with $(k_{\perp}^{\sigma},\sigma)$; $k_{\perp}^{\sigma}=0,\pi$. In the imbalanced case, the Fermi surfaces are mismatched,
and we  compute the pairing momenta ${\bf Q}=(Q_{\parallel},Q_{\perp})$
of possible FFLO-like states from the difference of the Fermi-wave vectors for majority and minority spins, {\it i.e.}, 
${\bf Q}={\bf k}_{F}^{\uparrow}-{\bf k}_{F}^{\downarrow}$,  
$Q_{\parallel}=k_{F\parallel}^{\uparrow}-k_{F\parallel}^{\downarrow}$, and 
$Q_{\perp}=k_{F\perp}^{\uparrow}-k_{F\perp}^{\downarrow}$.
Depending on $p$, $n$, and $t_{\perp}/t_{\parallel}$, 
the band structure can allow for more than one pairing wave vector $Q_{\parallel}$, yielding a quasi-condensate with 
multiple contributing modes. This may also be thought of as coexisting quasi-condensates with $Q_{\perp}=0$ and $Q_{\perp}=\pi$.
Examples for the expected polarization dependence
of the pairing momenta $Q_{\parallel}$  are displayed with lines in  
Fig.~\ref{fig:Q} for generic fillings of $n<1$ and  $t_{\perp}/t_{\parallel}=1$ and $2$.
To label these branches, we use a symbol $Q_{\perp}^{(k_{\perp}^{\uparrow},k_{\perp}^{\downarrow})}$, which specifies $Q_{\perp}$
as well as the $k_{\perp}^{\sigma}$ of the majority and the minority spins.
For instance, $\pi^{(\pi,0)}$ represents $Q_{\perp}=\pi$ pairs, formed from 
majority spins with $k_{\perp}^{\uparrow}=\pi$ and minority spins with $k_{\perp}^{\downarrow}=0$.

The disappearance of certain branches is due to a band getting completely filled or depleted, causing 
kinks in the other branches at the same $p$.
In the case of the ladder, there are two constraints on the four Fermi momenta, 
Luttinger's theorem for both spin flavors: $\sum_{k_{\perp}} k_{F,\parallel}^{\sigma}=2 \pi\,  \,N_{\sigma}/(2L)$ \cite{giamarchi,sachdev06}.
Thus, unlike the single-chain case where $n$ and $p$
determine the two Fermi momenta independently of $U$,  in the case of  a ladder,  interactions may modify the $Q_{\parallel}$ derived from the $U=0$ case.

\begin{figure}[t]
\includegraphics[width=0.44\textwidth]{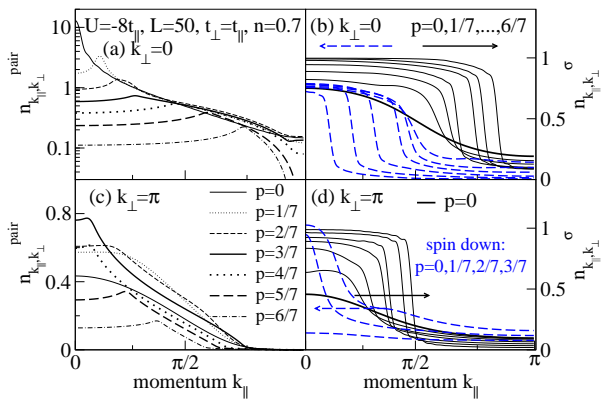}
\caption{(Color online) $t_{\perp}=t_{\parallel}$, $U=-8t_{\parallel}$, 
 and  $n=0.7$: (a), (c): Pair structure factor for (a): $k_{\perp}=0$ (c): $k_{\perp}=\pi$.
(b), (d):  MDF $n_{k_{\parallel},k_{\perp}}^{\sigma}$ ($\sigma=\uparrow$($\downarrow$): solid(dashed) lines).
 In all plots,  $p=0,1/7,\dots,6/7$ (except for $\sigma=\downarrow$ in (d); there, $p=0,1/7,2/7,3/7$).
 Arrows indicate increasing $p$.  The legend in (c) applies to (a),(c), the thick solid line in (b),(d) is for $p=0$.
}
\label{fig:pair_nk}
\end{figure}

Let us now come to the discussion of the (s-wave) pairing correlations at $|U|>0$. To that end, we compute the pair structure factor $n_{k_{\parallel},k_{\perp}}^{\mathrm{pair}}$
as a Fourier transform of $\rho_{l_1 l_2,rs}^{\mathrm{pair}}= \langle c^\dagger_{r,l_1,\uparrow}c^\dagger_{r,l_1,\downarrow} c_{s,l_2,\downarrow}c_{s,l_2,\uparrow} \rangle $, {\it i.e.}, 
\begin{equation}
n_{k_{\parallel},k_{\perp}}^{\mathrm{pair}} = \frac{1}{2L} \sum_{l_1,l_2} e^{ik_{\perp}(l_1-l_2)} \sum_{rs} e^{ik_{\parallel}(r-s)}\, \rho_{l_1 l_2,rs}^{\mathrm{pair}}\,. \label{nkpair}
\end{equation} 
This quantity is displayed in Figs.~\ref{fig:pair_nk}(a) and (c) for  $k_{\perp}=0,\pi$, respectively ($n=0.7$ and $t_{\perp}=t_{\parallel}$).
For $k_{\perp}=0$ [see Fig.~\ref{fig:pair_nk}(a)], there is a strong coherence peak at $Q_{\parallel}=0$ at $p=0$, which, upon polarizing, shifts to finite momenta $Q_{\parallel}>0$, as expected for
FFLO-like pairing.   In the $k_{\perp}=\pi$ channel, a  peak with $Q_{\parallel}>0$ emerges only for $p\gtrsim 0.2$, approaches $Q_{\parallel}=0$
at $p\approx 0.34$, and then $Q_{\parallel}$ increases again in the limit of large polarizations. 

We can further elucidate this behavior by analyzing the single-particle momentum distribution function (MDF)
$n_{k_{\parallel},k_{\perp}}^{\sigma}$, which is the Fourier transform of the one-particle density matrix $\rho_{l_1 l_2,ij}^{\sigma}=\langle  c^\dagger_{i,l_1,\sigma}
c^{}_{j,l_2,\sigma}\rangle$. We display the MDF, for the parameters of  Figs.~\ref{fig:pair_nk}(a) and (c), in Figs.~\ref{fig:pair_nk}(b) and (d).
Most notably, we see how the fraction of majority fermions increases for both  $k_{\perp}=0$ and $k_{\perp}=\pi$, with the $k_{\perp}=0$ channel dominating.
Secondly, as expected, the minority spins in the $k_{\perp}=\pi$ channel are depleted first at around 
$p\gtrsim 2/7$. Therefore, for instance, a $0^{(\pi,\pi)}$ quasi-condensate can only be realized at small $p$.

 In principle, the case of Fig.~\ref{fig:pair_nk} could allow for up to four combinations of $(Q_{\parallel},Q_{\perp})$
 at small $p$, as can be inferred from Fig.~\ref{fig:Q}(b). 
Yet, at $U<0$, we find at maximum one  peak in the pair structure factor in each channel.
We compare the position of these peaks  as extracted from the DMRG data
 to the $Q_{\parallel}$ computed in the $U=0$ limit, for the parameters of Fig.~\ref{fig:Q}.
 In the case of  $t_{\perp}=t_{\parallel}$, we choose the
  fillings  such that several cuts through phase IV of Fig.~\ref{fig:phase} can be followed (as indicated by the labels IVa-IVd in Fig.~\ref{fig:Q}).
We  subdivide phase IV according to how many 
bands are partially filled 
({\it i.e.}, $0<n<1$): IVa: four  bands; IVb: $(0,\sigma=\uparrow,\downarrow)$  partially filled  [$(\pi,\sigma=\uparrow,\downarrow)$ empty];
IVc: three bands [$(\pi,\downarrow)$ empty]; IVd: two bands [$(\pi,\uparrow)$ and $(0,\downarrow)$; $(\pi,\downarrow)$ empty; $(0,\uparrow)$ full].

At $n=0.7$ [Fig.~\ref{fig:Q}(b)], the $Q_{\perp}=0$ quasi-condensate at small $p$
 starts on the $0^{(\pi,\pi)}$ branch and then continously moves to the $0^{(0,0)}$ branch.  
At larger $p$,  a second condensate with $Q_{\perp}=\pi$ builds up  (circles), namely one of the $\pi^{(\pi,0)}$ type.
 Again, the positions $Q_{\parallel}$ of the quasi-condensates well agree with the $U=0$ predictions (lines in Fig.~\ref{fig:Q}) at polarizations $p\gtrsim 0.3$.
  In the case of a large filling $n=0.9$ and $t_{\perp}=t_{\parallel}$ [see Fig.~\ref{fig:Q}(c)], the picture is very similar
 to the $n=0.7$ case, with the difference that at large $p$, one enters into region IVd, 
where the $(0,\uparrow)$ band is filled, and hence only
 the $\pi^{(\pi,0)}$ branch survives.
  
 A simpler structure for the quasi-condensate is expected for the parameters of both Figs.~\ref{fig:Q}(a) and (d), where at maximum, two modes
 may exist. In these two cases
 all possible $(Q_{\parallel},Q_{\perp})$
 indeed contribute to the quasi-condensate at finite $U$ and, moreover, their momenta $Q_{\parallel}$ mostly agree with the $U=0$ predictions. 
An interesting effect occurs at 
 polarizations $p\gtrsim 0.35$ in Fig.~\ref{fig:Q}(d), where at $U=0$, no phase IVc is present. The interactions renormalize the dispersion such that 
the $(0,\uparrow)$ band
remains partially filled up to $p\lesssim 0.7$, allowing the $0^{(0,0)}$ branch to survive up to that value of polarization. 
Therefore, due to the interactions, a small window for the emergence of IVc opens up.
For $t_{\perp}>2t_{\parallel}$ and $n=0.7$, 
only one quasi-condensate is realized (not shown in the figures).
We emphasize that at any polarization and in all cases studied, at least one  mode with $Q_{\parallel}>0$ is present.

At small $p$,  the decay of 
pair correlations is consistent with  $|\rho_{ll,rs}^{\mathrm{pair}}| \sim |\cos(Q_{\parallel}|r-s|)|/|r-s|^{\alpha}$  
for $p\leq 2/7$, as expected for a single-${\bf Q}$ quasi-condensate \cite{yang01}. The same applies to all cases in which a single branch is present in
Fig.~\ref{fig:Q}. In all other cases, our data suggest that the envelope of $|\rho_{ll,rs}^{\mathrm{pair}}|$ follows a power-law.
 Moreover, we find that pair correlations decay slower than those of competing instabilities such as density-density correlations \cite{luescher08} at small $p$,
 while at present, we cannot make a definite statement about larger polarizations.
 Out of the possible quasi-condensates, typically the ones with the {\it smallest} $Q_{\parallel}$ win and show up 
 with a significant weight in the pair structure factor. The most 1D-like behavior with a single $(Q_{\parallel}, Q_{\perp}=0)$ mode is encountered in phase IVb, which for 
 $t_{\perp}/t_{\parallel}=1,2$ extends up to full polarization for $n\lesssim 0.4$. In IVb, $Q_{\parallel}=2\pi n\,p$, as in 1D \cite{feiguin07}.

\begin{figure}[t]
\includegraphics[width=0.44\textwidth]{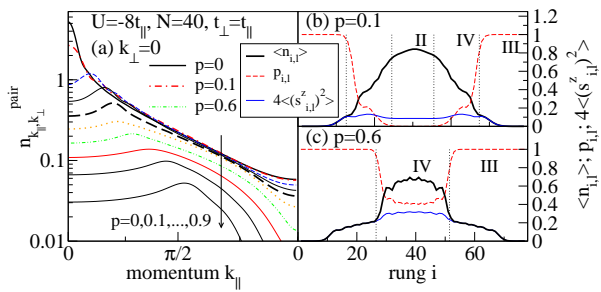}
\caption{(Color online) Harmonic trap $V_{\mathrm{trap}}=0.002 t_{\parallel}$; 
$t_{\parallel}=t_{\perp}$, $N=40$, and  $U=-8t_{\parallel}$.
(a)
Pair structure factor for $k_{\perp}=0$ and $p=0,0.1,\dots,0.9$.
(b), (c): Density profile $\langle n_{i,l} \rangle$ ($n_{i,l}=n_{i,l,\uparrow}+n_{i,l,\downarrow}$), local polarization $p_{i,l}$, and local  
$\langle (s_{i,l}^z)^2\rangle =\langle (n_{i,l,\uparrow}-n_{i,l,\downarrow})^2 \rangle/4$   for (b) $p=0.1$ and (c) $p=0.6$.
Vertical, dotted lines are a guide to the eye.}
\label{fig:trap}
\end{figure}

We now turn to the effect of a harmonic trapping potential on the pairing correlations and the density profile
by setting $V_{\mathrm{trap}}=0.002t_{\parallel}$ with $N=40$ fermions and $t_{\perp}=t_{\parallel}$.
As for phase separation, our results displayed in Fig.~\ref{fig:trap} indicate that at small polarizations ($p< 0.2$), there are three
shells: an inner core with a vanishing polarization, a thin shell that is partially polarized with increased spin fluctuations, 
and fully polarized wings [see Fig.~\ref{fig:trap}(b), $p=0.1$ there]. 
This is distinctly different from the case of uncoupled chains with the same $U$ and filling \cite{feiguin07,hu07}, where FFLO-like correlations develop
in the core as soon as $p>0$. 
By invoking the local density approximation, we trace this back to the  slope of the boundary between phases II and IV that has a different sign 
in the case of ladders (see Fig.~\ref{fig:phase}) as compared to chains \cite{essler-book}; thus the system goes through IV in the center of the trap before entering III
toward the edges.
At $U=-8t_{\parallel}$, the formation of fully polarized wings sets in as soon as $p>0$.

 At larger polarizations, there are only two phases: in the core, the local polarization $ p_{i,l}= (n_{i,l,\uparrow}-n_{i,l,\downarrow})/(n_{i,l,\uparrow}+n_{i,l,\downarrow})$ increases and the unpolarized phase gives room to a partially polarized one, while the wings
 remain fully polarized [see Fig.~\ref{fig:trap}(c) for the example of $p=0.6$]. We relate the phases emerging in the trapped situation to those of the bulk system shown in
 Fig.~\ref{fig:phase}: the unpolarized phase is a $Q_{\parallel}=0$ superfluid -- phase II -- while the partially polarized one has, similar to phase IV, FFLO-like features. 
 Our results for the pair structure factor, presented in Fig.~\ref{fig:trap}(a), confirm this picture: For $p<0.2$, a coherence peak at $Q_{\parallel}=0$ dominates
 in the $k_{\perp}=0$ channel, which, as soon as the core assumes a finite polarization, develops into a $Q_{\parallel}>0$ peak, characteristic of an FFLO state.
 The evolution of the coherence peaks for both  $k_{\perp}=0$ and $\pi$  resembles that of the untrapped system [Figs.~\ref{fig:Q}(a) and (b)].
 The coherence peaks at both $Q_{\perp}=0$ and $\pi$ survive up to saturation.
 At high densities (not shown here), a band-insulating Fock state with uncorrelated tightly bound pairs appears in the center of the trap, displacing the other phases toward the edges,  
 similar to the 1D \cite{feiguin07} and the 2D cases \cite{iskin08}.

\emph{Discussion --}
The physics of ladders already features some characteristics of higher dimensional systems. In particular,
 the positive slope of the boundary of phase II (see Fig.~\ref{fig:phase}) is a remarkable  effect of the dimensionality 
 that  changes the order of the concentric phase-separated shells in a trap, in agreement with mean-field calculations in 2D and 3D.
To learn about the incipient 2D physics, it is edifying to look
 at a ladder with
$t_\perp=2t_\parallel$. 
In this case, the IVb and IVd regions occupy most of  phase IV,  
with only a small sliver of IVc at high densities due to the renormalization of the bandwidth [see Fig.~\ref{fig:Q}(d)]. 
In the isotropic 2D system, region IVb would correspond to two
partially filled bands, while IVd has  a band of majority spins
above half filling, touching the boundaries of the Brillouin zone.
At a low density (phase IVb), the physics fits within a
single-band picture [see, {\it e.g.}, Fig.~\ref{fig:Q}(a)], 
similar to a   chain.
At larger densities, the  pairing acquires
contributions from multiple Fermi points.
Similarly, in higher dimensions and
at low density, the problem has rotational symmetry,
and the FFLO order parameter can be faithfully described by the modulus
of a single wave vector $|\bf{Q}|$.
However, at higher densities the Fermi surface(s) acquires a diamond-like
shape, and a multi-modal description is expected to be more accurate \cite{bowers02}.
This would translate into complex real-space modulations of the order
parameter, such as those observed in Ref.~\cite{iskin08}.
Beyond 1D, nesting between bands 
with opposite spin is weak \cite{3D-imbalanced}, particularly at high densities and high magnetic 
fields, when the Fermi surfaces have very different shapes, with a 
large mismatch in Fermi velocities. Thus, the FFLO state 
becomes unfavorable compared to a normal polarized state.
The ladder system, however, is strongly nested to the effect that
we find a pairing instability of the FFLO-type at all $p$ and $n$ studied, with a much richer structure
than in the single-chain case.

We thank D.~Huse and G.~Roux for fruitful discussions.


\end{document}